\documentclass[a4paper,11pt]{article}
\usepackage[svgnames]{xcolor}
\usepackage{moreverb}
\usepackage{graphicx,amssymb,amstext,amsmath,tikz}
\usepackage[para]{threeparttable}
\usepackage{mathtools,xparse}
\usepackage{natbib}
\usepackage{bbold}

\usetikzlibrary{positioning,shapes.geometric}
\usepackage{rotating}
\usepackage[T1]{fontenc}
\usepackage{longtable}

\usepackage{float}
\usepackage[caption = false]{subfig}
\usepackage{natbib}

\usepackage[left=2cm,top=2cm,right=2cm, bottom=2.5cm]{geometry}
\usepackage{setspace}
\usepackage{enumitem}
\usepackage{pdfpages}
\usepackage{amssymb}
\def\bSig\mathbf{\Sigma}

\usepackage{moreverb}
\usepackage{graphicx,tikz}
\usepackage[para]{threeparttable}
\usetikzlibrary{positioning,shapes.geometric}
\usepackage{rotating}
\usepackage{longtable}

\definecolor{mygray}{gray}{0.67}

\usepackage{enumitem}
\def\s{\sigma^2}

\def\logit{\mbox{logit}}

\def\ci{\perp\!\!\!\perp}

\begin{document}

\title{Local average treatment effects estimation via substantive model compatible  multiple imputation}

\author{K.~DiazOrdaz and J.~ Carpenter\\\footnotesize{Department of Medical Statistics,} \\\footnotesize{London School of Hygiene and Tropical Medicine, U.K. }
}

\date{}
\maketitle


\begin{abstract}
Non-adherence to assigned treatment is common in randomised controlled trials (RCTs). Recently, there has been an increased interest in estimating causal effects of treatment received, for example the so-called local average treatment effect (LATE). Instrumental variables (IV) methods can be used for identification, with estimation proceeding either via fully parametric mixture models or two-stage least squares (TSLS). TSLS is popular but can be problematic for binary outcomes where the estimand of interest is a causal odds ratio. Mixture models are rarely used in practice, perhaps because of their perceived complexity and need for specialist software.
Here, we propose using multiple imputation (MI) to impute the latent compliance class appearing in the mixture models. Since such models include an interaction term between  compliance class and randomised treatment, we use ``substantive model compatible'' MI (SMC MIC), which can also address other missing data, before fitting the mixture models via maximum likelihood to the MI datasets and combining results via Rubin's rules.
We use simulations to compare the performance of SMC MIC to existing approaches and also illustrate the methods by re-analysing a RCT in UK primary health.
We show that SMC MIC can be more efficient than full Bayesian estimation when auxiliary variables are incorporated, and is superior to two-stage methods, especially for binary outcomes.


\end{abstract}

\textbf{keywords:}
Instrument variables, local average treatment effect, missing data, multiple imputation, non-adherence.

\section{Introduction}

Many empirical studies are concerned with  estimating a causal effect of an  exposure on one or more outcomes of interest. Ideally, this type of question should be answered in a randomised controlled trial (RCT).  
Nevertheless,  randomisation to certain exposures is not always possible,  and even when it is, RCTs often suffer from non-compliance with allocated treatment \citep{Dodd2012}. 
In these situations, if the exposure or treatment received  is confounded, in the sense that there are measured and unmeasured common causes of the exposure and the outcome,  then inferences based on estimators that fail to adjust for this will  be invalid \citep{White2005}. 

In the presence of unmeasured confounding, instrumental variable (IV) methods can be used  to estimate consistent causal effects \citep{Imbens1994,  Angrist1996, Frangakis2002}. An IV is a variable which is correlated with the exposure but  is not associated with any confounder of the exposure--outcome association, nor is there any pathway by which the IV affects the outcome, other than through the exposure.  

Once we have identified  a suitable IV then, depending on the additional assumptions we are prepared to make,  different causal treatment effects can be point-identified and  estimated. Here, we focus on the local average treatment effect (LATE),  also referred to as complier-average causal effect (CACE)\citep{Angrist1996}.

For estimation, two popular approaches are mixture models and standard econometric IV techniques, such as  two-stage least squares estimator (TSLS) \citep{Angrist1995}. 
The first approach uses a fully-parametric structural model for the outcome, depending on latent compliance class membership, the IV and potentially other baseline covariates. Then,  Bayesian \citep{Imbens1997, Frangakis2002} or maximum likelihood \citep{Yau2001, Jo2001} estimation can be used.
The second appoach, TSLS, consists of a ``first stage'', which regresses the exposure on the IV, while the second stage consists of regressing the outcome on the predicted exposure, coming  from the first stage regression. The coefficient corresponding to the predicted exposure in this second model is  the TSLS estimator of the desired causal treatment effect. TSLS is a consistent estimator as long as both  stages are linear regressions \citep{Wooldridge2010}. 
This is an issue in settings with binary outcomes where interest may be in estimating causal odds ratios. Two-stage  (TS) estimators which use nonlinear regressions have been proposed (See \citep{Didelez2010, Clarke2012} for reviews) but, due to the non-collapsibility of the odds ratios,  these are not necessarily consistent \citep{Cai2011}.

The mixture models do not suffer from this, but in general require specialist software to be fitted. 
To address this,  we propose a new method for estimating the CACE based on Multiple Imputation (MI)  \citep{Rubin1987}. MI is a practical and flexible method widely used  to handle missing data.  MI has been proven to be asymptotically equivalent to Bayesian estimation under certain conditions \citep{Liu2013}, so our proposal should provide equivalent estimates with good frequentist properties. Compared with standard IV estimation methods, MI estimation should have increased precision, and has the added benefit of seamlessly handling the missing data in outcomes or covariates.  

The rest of the paper proceeds as follows. Section \ref{Sec:PS} formally introduces the CACE and presents the assumptions necessary for identification. The proposed methods are developed in Section \ref{Sec:MI-C}. Section \ref{Sec:Simulation} presents simulation studies demonstrating the empirical performance of the proposed method in finite sample settings, and comparing it to that of Bayesian and two-stage methods. In Section \ref{Sec:COPERS}, we illustrate the methods by re-analysing the COPERS (COping with persistent Pain, Effectiveness Research in Self-management) trial, a UK based RCT testing a cognitive behavioural therapy intervention designed to help managing chronic back pain.  Section \ref{Sec:Discussion} concludes with a discussion.

\section{Local average treatment effects}\label{Sec:PS}

For the remainder of the paper,  we consider the  setting of a RCT with non-compliance, but translation to other settings with a valid IV is immediate. We use ``treatment'' to represent either treatment received or exposure to a risk factor, while the IV is represented by randomised treatment.

Consider  a  two-arm controlled randomised clinical trial, with $N$  randomised participants. Let $Z$ and $D$ denote the treatment randomly allocated and received respectively, both assumed to be binary, while the outcome $Y$ can be continuous or binary.  Let $U$ be the set of all unobserved common causes of $D$ and $Y$, that is the unmeasured confounders.
For simplicity, we assume that the active treatment is subject to all-or-none time-invariant compliance, and that the control group does not have access to the active intervention. 

Let $D(z)$ be the potential treatment received corresponding to the random treatment allocation $z$.  Similarly, let $Y(z,d)$ be the potential outcome under random allocation $z$ and receiving treatment $d$. 

Throughout, we assume the  \textbf{Stable Unit Treatment Value Assumption (SUTVA)} holds,
which comprises  \textit{no interference}, i.e. the potential outcomes of the $i$-th individual are unrelated to the treatment status of all other individuals, and \textit{consistency},  for all individuals $i=1,\ldots, N$, if $Z_i=z$ then $Y_i(z, D(z))=Y_i$, for all $z$.  

 In the setting where both $Z$ and $D$ are binary, the vector of potential treatment received under alternative random allocation, $(D_i(0), D_i(1))$ partitions the population into four different \textit{compliance classes}: $C_i = n$ (never-taker), if $D_i(0) = D_i(1) = 0$; $C_i = a$ (always-taker), if $D_i(0)= D_i(1) = 1$; $C_i = c$ (complier) if $D_i(z) = z$ for $z\in\{0,1\}$; and $C_i = d$ (defier) if $D_i(z) = 1-z$ for $z\in\{0,1\}$. These latent classes are unaffected by the realised $Z$, so they can be treated as baseline variables in the analyses. 

\subsection{Estimands, assumptions and identification}\label{Sec:IVassumptions}

Under SUTVA, we can define the average causal treatment effect  in the compliers,  the CACE, as
$E\left[\{Y_{i}(1,D(1))-Y_{i}(0,D(0))\}\big|\{D_i(1)-D_i(0)=1\}\right],$
or equivalently $\mbox{CACE} = E[Y_i(z=1) - Y_i(z=0)\big|C_i = c].$
This is said to be a ``local'' effect as it is conditional on belonging to the compliers stratum.

For identification, in addition SUTVA, the following core IV assumptions are often made \citep{Angrist1996}: \textbf{(A1)}  \textbf{Unconfoundedness} : $Z\ci U$, 
\noindent\textbf{(A2)}  \textbf{exclusion restriction}: 
 $Z \ci Y \vert D, U$,  i.e. conditional on the treatment received and the unobserved confounder $U$, the instrument and the outcome are independent.
This is often explained as the effect of $Z$ on $Y$ must be via an effect of $Z$ on $D$; $Z$ cannot affect $Y$ directly. 
\noindent\textbf{(A3)} \textbf{Instrument relevance:} Also referred to as first stage assumption: $Z$ is causally associated with treatment received $D$, i.e. $Z \not\perp\!\!\!\perp D$.

Finally, for point identification of the LATE,  \textbf{(A4) monotonicity} of the treatment mechanism is often assumed,  $D(1)\geq D(0)$, often informally referred to as ``there is no defiers'' \citep{Imbens1994}.

 In the RCT settings considered here, where the individuals randomised to control have no access to the experimental treatment, a stronger from of monotonicity, often called ``no contamination'', is justified by design.
In this special case,  there is only two compliance strata: compliers, and never-takers.

Under these assumptions (SUTVA and A1-A4), and in  settings without covariate adjustment, the marginal CACE is identified from the observed data by the Wald (or ratio) estimand \citep{Imbens1994}
\begin{equation}\label{Wald}
\beta_{IV} = \frac{E(Y|Z=1) - E(Y|Z=0)}{E(D|Z=1) - E(D|Z=0)}
\end{equation}
Alternatively,  the CACE is identified by the intention-to-treat (ITT) estimand of the IV on the outcome of interest in the sub-population of compliers \citep{Imbens1997}.   \citet{Angrist1996} showed that the Wald estimand (eq. \ref{Wald}) is in fact the ITT effect in the compliers.

In settings where outcomes are missing, it is important to condition on compliance class, before assuming that the missingness  mechanism  is independent of the outcome. More formally, let $R$ be the missingness indicator for $Y$,  equal to 1 if the outcome is observed, and 0 otherwise, we assume \textbf{(A5)} \textbf{latent ignorability}: within each latent compliance class, the missing data is independent of the outcome, given the IV and baseline covariates (if conditioning on any) \citep{Yau2001, Taylor2009}: $P[R_i(z)\vert Y_i(z),C_i] = P[R_i(z)\vert C_i], \quad z\in\{0,1\}.$

We also need to replace assumption (A2) with:
\textbf{(A2') compound exclusion restriction}: 
conditional on the exposure and the unobserved confounder, the level of the IV does not affect the outcomes or missingness mechanism, i.e.:
$Z \ci Y , R \vert D, U$.
This is often expressed  as $P[Y_i(1),R_i(1)\vert C_i ] = P[Y_i(0),R_i(0)\vert C_i ]$.

\section{Multiple Imputation of compliance classes and estimation of mixture models} \label{Sec:MI-C}
In the  mixture model approach, estimation proceeds by specifying a fully parametric
model for the outcome, the partially latent compliance classes and the IV (and  baseline covariates if using any) \citep{Imbens1997, Little1998, Hirano2000}.

Here we are considering continuous or binary outcomes,  so we assume  $Y_i\sim N(\eta_Y,\s_Y)$ or $Y_i\sim\mbox{Bern}(\eta_Y)$  respectively.  Under ``no contamination'', there is no always-takers and no defiers, so the compliance class is binary  $C_i\sim\mbox{Bern}(\pi)$, with $\pi$ independent from $Z$.  
The parametric model used to estimate the marginal LATE is an extended general location model \citep{Yau2001}, with the following form 
\begin{equation}\label{modY}
g(\eta_Y)=\beta_0 +\beta_c C + \beta_{cz} CZ 
\end{equation}
where $g$ is the identity or logit link function, corresponding to whether the outcome is  continuous or binary. The CACE is estimated by the coefficient of the interaction term, $\widehat{\beta}_{cz}$.  Note that because of the exclusion restriction, there is no direct effect of $Z$ on $Y$.
We call this the \textit{analysis or substantive} model.

 Inference based on this model can be more efficient than standard IV methods \citep{Little1998}, but is sensitive to unverifiable parametric assumptions \citep{Tan2006}. 

Estimation of the parameters in model \ref{modY}  is usually done via maximum likelihood, using Expectation-Maximisation (EM)  \citep{Little1998, Jo2002} or Bayesian estimation \citep{Imbens1997,Yau2001}.

Here, we propose to frame the problem as one of missing compliance data, so that after imputing these, estimation of the substantive model can proceed by maximum likelihood, and the estimates then combined using Rubin's rules \citep{Rubin1987}.
For MI to result in valid inferences,  the imputation model must include all the terms that appear in the substantive model. The problem for the mixture model used for CACE, eq. \ref{modY}, is that it contains an interaction term involving the compliance class, which needs to be imputed. Imputation of interaction terms is complex. The short-comings of ad-hoc alternatives (e.g. passive imputation) are well documented  \citep{Seaman2012}. 
The key to produce correctly specified imputations is to use MI routines which allow for substantive-model compatible imputation. One such  method is the so-called substantive model compatible full conditional specification, or SMCFCS \citep{Bartlett2014}.

The generic SMCFCS procedure works as follows. Suppose we have  partially observed covariates $X_{1},X_{2},..,X_{p}$, and fully observed covariates $\mathbf W$. Denote the distribution of  $Y$ implied by the  substantive model  by $[Y|X_{1},..,X_{p},\mathbf Z,\psi]$, with parameters $\psi$. We must impute   the $j$-th  partially observed covariate $X_j$,  from an imputation model  for $[X_{j}|X_{-j},\mathbf W,Y]$ which is compatible with this.  A compatible imputation model for $X_j$    can be expressed as
	\begin{equation}
[X_{j}|X_{-j},\mathbf W,Y]= \frac{[Y, X_{j},X_{-j},\mathbf W]}{[Y, X_{-j},\mathbf W]} 
\propto 	[Y|X_{j},X_{-j},\mathbf W][X_{j}|X_{-j},\mathbf W].
	\end{equation}
	
We can thus specify an imputation model for $X_{j}$ which is compatible with the substantive model by additionally specifying a model for $[X_{j}|X_{-j},\mathbf W]$.
With categorical outcomes, this has a closed expression, but  in general the implied imputation model will not be from a standard distribution, 
and rejection sampling is used to obtain draws.
	
The SMCFCS algorithm initialises by imputing missing values in each variable using randomly observed values from the same variable. It then cycles through the imputation models for each partially observed variable,  imputing each in turn.  This needs to be iterated  a number of times, so that the draws are taken from the (approximate) full conditional distributions after they have converged to the stationary distributions. The process is then repeated to create $M$  imputed datasets, which are then analysed with the substantive model, and combined with Rubin's rules, as per usual. The estimates obtained in this way are very  close to those obtained with a full Bayesian analysis with similarly vague priors, as full conditional specification (FCS) MI has been shown to be equivalent to full Bayesian analysis where the sequential conditional models are compatible  \citep{Liu2013}.

Where there are missing values in the outcome,  it may be necessary to condition on baseline covariates in order for $R$ to be latent ignorable (i.e. $Y$ is missing at random (MAR) given the compliance class, the IV and other observed variables).   If these variables are not included in the analysis model,  they should  be  included in the imputation model.

For the mixture models used for CACE estimation, we need to impute the partially missing compliance class $C$, in a way that accommodates the interaction term $CZ$, whose coefficient represents the CACE.  Applying the SMCFCS principle means that we need to specify an imputation model   $[C\vert Z]$, (and any auxiliary variables if using) as well as specifying the substantive model (eq. \ref{modY}).
The smcfcs R and Stata package then performs the imputation of $C$ given $Z$, and rejects the proposed value if it is not compatible with the substantive model. Sample R code specifying the imputation and substantive models can be found in the Web Appendix A. We denote this method SMC MIC (substantive model compatible multiple imputation of compliance).
 
\subsection{LATE estimation by standard  IV methods}

Estimation of the conditional expectations appearing in the Wald estimand  $\beta_{IV}$ (eq. \ref{Wald}) 
is often done by  two-stage least squares (TSLS). The first stage fits a linear regression to treatment received on treatment assigned, $ D_i = \alpha_0 + \alpha_1Z_i + \omega_{1i}$. 
Then, in a second stage,  a regression model for the outcome on the predicted treatment received is fitted, $ Y_{i} = \beta_0 + \beta_{IV} \widehat{D_i} + \omega_{2i}$,  where  $\omega_{1},  \omega_{2}$ are the residuals which are not assumed to follow a normal distribution. Covariates can be used by including them in both stages of the model.  The asymptotic standard error for the TSLS is given in \citep{Imbens1994}, and implemented in commonly used software packages.

Crucially both first and second stages must be linear models for the TSLS estimator to be consistent \citep{Wooldridge2010}. Thus, for binary outcomes, the  TSLS  is a consistent estimator for the local risk difference, but it imposes strong assumptions on probability bounds and constant effects of exposure to the IV \citep{Imbens2001}.  

However,  where the estimand of interest is the local odds ratio,  several estimators have been proposed, all exhibiting a smaller or greater degree of bias under certain circumstances. The logistic Wald-type estimator \citep{Palmer2008}
\begin{equation} 
\mbox{WaldOR} =\mbox{exp}\left[\frac{\mbox{logit}\{E(Y \vert Z = 1)\} - \mbox{logit}\{E(Y |Z = 0)\}}
{E(D|Z = 1) − E(D|Z = 0)}\right]
\end{equation}
has been  shown to be a good approximation to the local odds ratio when $D$ is symmetrically distributed, with the bias increasing  with increasing values of the true causal association between $D$ and $Y$ given $Z$, and with increasing residual variance in $D$ given $Z$ \citep{Vansteelandt2011, Clarke2012}.  Thus, we shall not consider this any further.

Two-stage (TS) estimators with non-linear first or second stages are not in general consistent under misspecification of either model, and therefore are not  recommended. Nevertheless, we consider here  the plug-in  residual inclusion (TSRI) method  \citep{Terza2008}.  The first stage is similar to that of the TSLS,  but instead of using the fitted values for $D$, we calculate the fitted residual, $\widehat{V} = D - \widehat{\alpha}_0 + \widehat{\alpha}_1Z$.  In stage two, we fit  $\mbox{logit}(Pr(Y=1)= \beta_0 + \beta_1 D + \beta_2\hat{V}$.  Now the estimated coefficient on $D$ is an  estimator for the CACE. The asymptotic standard error for the TSRI CACE is given in  \cite{Terza2014}.

This approach is easy to implement, but in order to be consistent it requires that the mean potential outcome in the unexposed never-takers is equal to the mean potential outcome of the unexposed compliers, $\eta_{n0}=\eta_{c0}$ \citep{Cai2011}.  This is because an IV analysis adjusted for the residuals of the first stage is equivalent to adjusting for the unmeasured confounders, and due to the non-collapsibility of the odds ratios, this will not in general correspond to the marginal causal odds ratio of interest \citep{Burgess2015}. 

In settings with missing data,  in addition to A1-A4, we assume that the missingness is conditionally independent of the outcomes given the covariates in the model  \citep{White2010}. Since IV models traditionally use no covariates,  it may be preferred to assume the missing data are  MAR given some observed covariates, and use these in a MI procedure prior to performing TSLS or TSRI.

\section{Simulation study} \label{Sec:Simulation}
We now perform  a simulation study comparing the finite sample performance of  (i) SMC MIC and (ii) Bayesian estimation of the mixture model (eq. \ref{modY}), and (iii) either TSLS or TSRI, for continuous and binary outcomes respectively. 

We simulate clinical trial data with one-way non-compliance to randomised  treatment, varying the proportion of non-compliers, the  sample size, the true value of the CACE and the type of outcome, as well as whether this is fully observed, or not. The factorial design is summarised in Table \ref{simulation_factors}.

 Two thousand datasets per scenario are obtained with the following the data generating mechanisms. We begin by simulating bivariate normal baseline covariates $X_1$ and $X_2$, with mean zero and covariance matrix  $\Sigma= \left(\!\!\!\begin{array}{cc} 1 & 0.3 \\ 0.3 & 1\end{array}\!\!\!\right)$, and randomised treatment $Z_i\sim \mbox{Bern}(0.5)$. Only $X_1$ is a true confounder, it is simultaneously associated with both treatment received  (and therefore compliance) and outcome. The second covariate $X_2$ is conditionally independent from the compliance class given $X_1$, but is associated with the outcome.  The confounder $X_1$ is assumed to be unmeasured, while $X_2$ is fully observed in all settings. Where the outcome is only partially observed, it is assumed MAR given $X_2$.
 
The precise parametric data generating model is as follows. We simulate the binary compliance class (assuming no always-takers or defiers) 
\begin{equation}\label{modC_dgp}
C \sim \mbox{Bern}(\pi), \quad 
\mbox{logit}(\pi)= \psi_0 +\psi_{x_1}X_1
\end{equation}

The outcome $Y\sim N(\eta_Y,1)$ or $Y\sim \mbox{Bern}(\eta_Y)$, with 
\begin{equation}\label{modY_dgp}
g(\eta_{Y})= \beta_0+ \beta_c C + \beta_{cz} CZ  +\beta_{x_1} X_1 +\beta_{x_2} X_{2}
\end{equation}
where $g$ is identity or logit link function, respectively.  

We consider two values for $\psi_0$, namely $0.85$ and $0.5$, so that the expected proportion of compliers is 0.7 and 0.5 respectively, chosen as a systematic review \citep{Dodd2012}  found that the percentage of non-compliance was less than 30\% in two-thirds of published RCTs, but greater than 50\% in one-tenth of studies. 
The true causal conditional  CACE in the link scale is represented by $\beta_{cz}$, and  takes  two values, small (=2) and large (=4). 
The rest of the parameters in the outcome model (eq. \ref{modY_dgp}) are fixed, with $\beta_0=0$, $\beta_{x_1}=-2.2$ and $\beta_{x_2}=0.5$. 

Given that TSRI is known to be consistent where $\eta_{n0}=\eta_{c0}$, we choose  $\beta_c=0$ in a first set of simulations. To compare the methods' behaviour on settings where TSRI is known to be biased, a second simulation set is performed only for binary outcomes settings, with $\beta_c=\frac{\beta_{cz}}{2}$.

Finally, the missingness mechanism  for $Y$ is $\logit(P(R)=1)= -1.386294 + \log 2 X_2$, so that on average 20\% outcomes are MAR given $X_2$.

For binary outcomes, given that the data are generated under a conditional model, the marginal CACE log odds ratio  is empirically calculated using  a dataset of size  $N=10,000,000$. 

All analyses are performed in R. The SMC MIC is implemented using the package smcfcs, with the number of iterations set to  250. For continuous outcomes, a maximum number of attempts made for the rejection sampling step is set to  5,000. These values are chosen by examining the  trace plots of a few randomly selected datasets, to study the convergence behaviour. The number of multiply imputed datasets is $M=10$.
For scenarios with missing outcomes, we simply allow SMCFCS to impute these as well, including $X_2$ as an auxiliary covariate in the imputation models. 

Bayesian methods are run in JAGS from R using the r2jags package. The median of the posterior distribution is used as the point estimate of the parameter of interest, and the standard deviation of the posterior distribution as the standard error. Equi-tailed 95\% posterior credible intervals are obtained, and henceforth referred to as confidence intervals (CIs), to have a unified terminology for both Bayesian and frequentist intervals.
 Two chains, each one with 10,000 initial iterations and 5,000 burn-in are used. 
For all the regression parameters appearing in the models,  a vague prior $N(0,0.001)$ (parameterised by the precision) is used, with the exception of scenarios with binary outcomes, where a vaguely informative prior  $N(0,0.02)$ is used for the log odds parameters, given that log odds ratios are rarely larger than 10 \citep{Gelman2008}. For continuous outcomes, the  prior on the standard deviation is $\sigma \sim \mbox{Gamma}(0.01,0.01)$. The prior for the probability of being a complier $\pi$ is $\mbox{Beta}(1,1)$.
Missing outcomes are accommodated by  adding a line to the model, including $X_2$ as an auxiliary variable. 

The TSLS method for continuous outcomes  was implemented using the ivreg command from the R package AER. The code implementing TSRI can be found in the Web Appendix B.  For scenarios where the outcome is missing, we perform MI prior to any TS analysis, using the R package mice, including treatment received and $X_2$ as an auxiliary variable in the imputation model, and run separately by the groups defined by $Z$, with $M=10$.  The  TS estimates obtained on each multiply imputed dataset are combined using Rubin's rules. 

After analysing each dataset with either full Bayesian, SMC MIC, or TS method,  we obtain the bias and its Monte Carlo error CI (MCE CI), coverage and width of the 95\% CIs, and root mean square error (RMSE). IV methods are known to be biased in finite samples, with the bias being small in strong IV settings. Thus bias smaller than 5\%  is considered acceptable \citep{AngristPischke2009}. In terms of coverage of the 95\% CIs, a method is usually considered  as underperforming  if its coverage drops below 90\% \citep{Collins2001}. Coversely, if coverage is very close to 100\%,  extra caution with the methods is required \citep{Yucel2010}.

\subsection{Simulation results}

\begin{figure}
\caption{Estimated CACE for continuous  outcome: performance (bias with MCE CI, coverage rates, CI width and RMSE) of  Bayesian, SMC MIC and TSLS methods.  The scenarios' sample sizes  and the \% of noncompliers varies by column.  The boxes in the plots show the following: \textcolor{mygray}{$\blacksquare$}  ``small'' true CACE and fully observed outcome, $\blacksquare$ ``large'' true CACE and fully observed outcome,  \textcolor{mygray}{$\square$} ``small'' true CACE but partially observed outcome, and $\square$ ``large'' true CACE but partially observed outcome. The dotted line in the bias plot represents no bias. Dashed lines on the coverage plot are the 94 and 96 \% coverage rates.}
\label{fig:cont}
\begin{center}
\includegraphics[scale=.9]{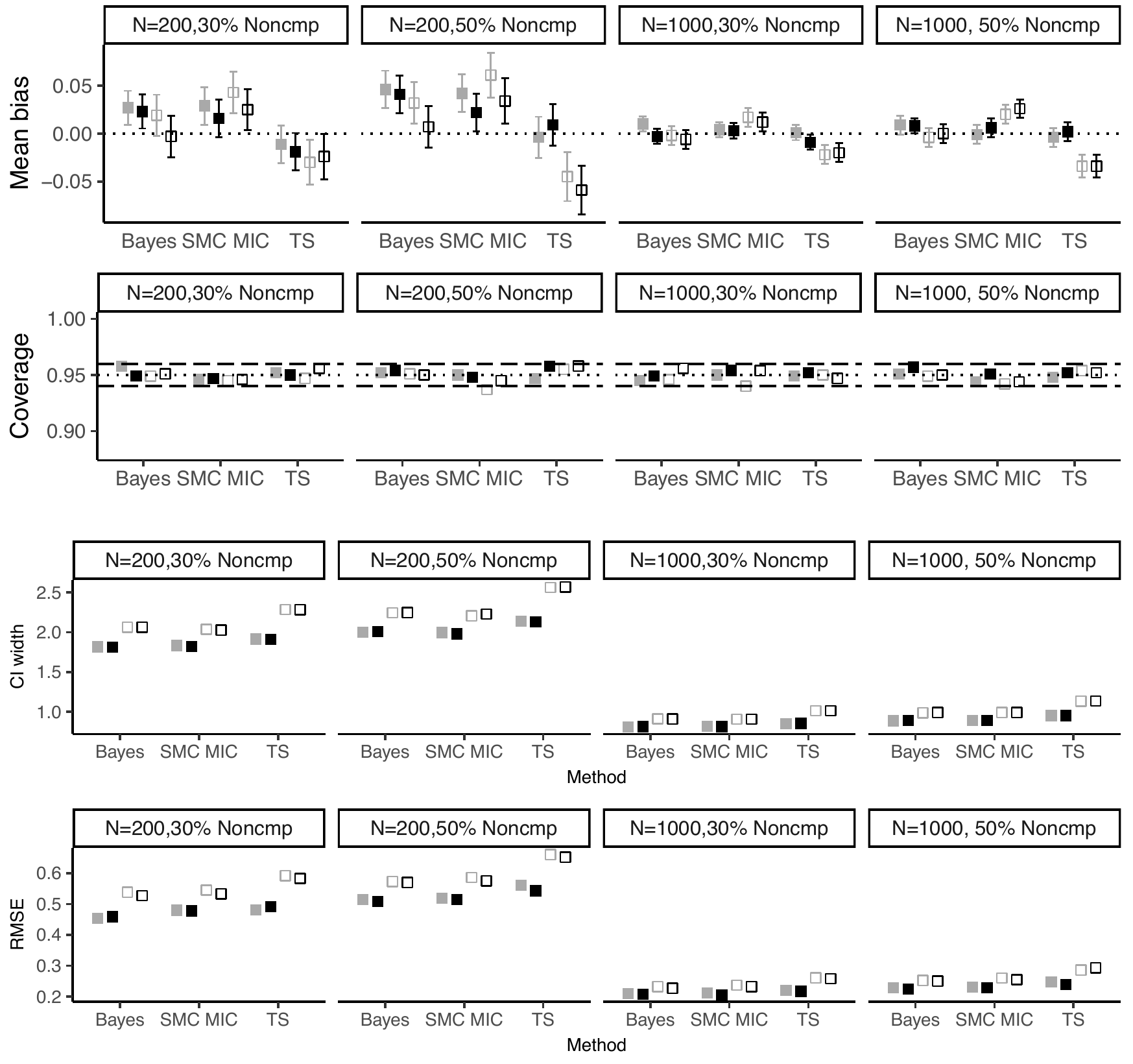}
\end{center}
\end{figure}

\begin{figure}
\caption{Estimated CACE for binary outcome where there is no effect of complying in the control arm: performance (bias with MCE CI, coverage rates, CI width and RMSE) of  Bayesian, SMC MIC and TSLS methods.  The scenarios' sample sizes  and the \% of noncompliers varies by column.  The boxes in the plots show the following: \textcolor{mygray}{$\blacksquare$}  ``small'' true CACE and fully observed outcome, $\blacksquare$ ``large'' true CACE and fully observed outcome,  \textcolor{mygray}{$\square$} ``small'' true CACE but partially observed outcome, and $\square$ ``large'' true CACE but partially observed outcome. The dotted line in the bias plot represents no bias. Dashed lines on the coverage plot are the 94 and 96 \% coverage rates.}
\label{fig:binary}
\begin{center}
\includegraphics[scale=.9]{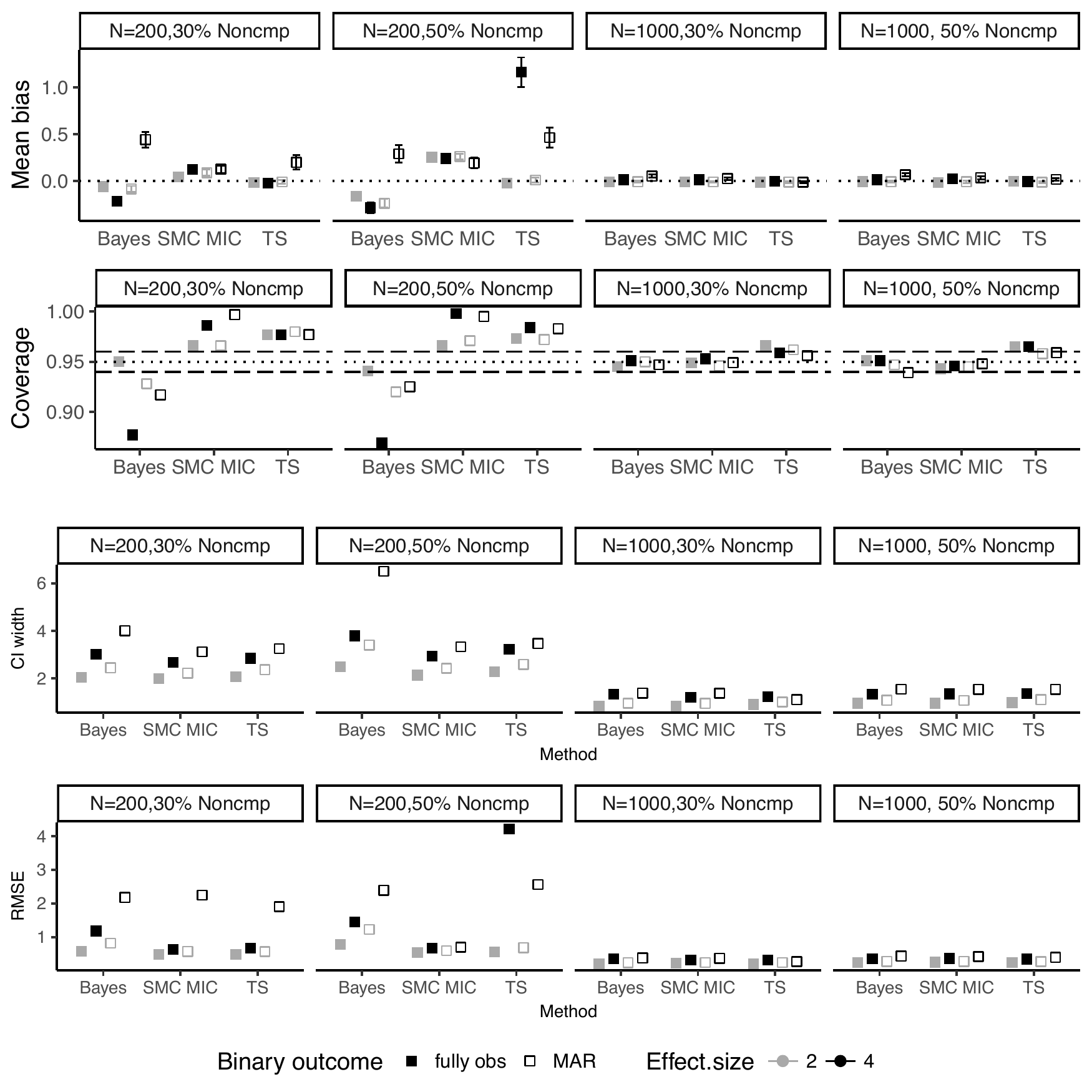}
\end{center}
\end{figure}

\begin{figure}
\caption{Estimated CACE for binary outcome  non-zero effect of complying in the control arm: performance (bias with MCE CI, coverage rates, CI width and RMSE) of  Bayesian, SMC MIC and TSLS methods.   The scenarios' sample sizes  and the \% of noncompliers varies by column.  The boxes in the plots show the following: \textcolor{mygray}{$\blacksquare$}  ``small'' true CACE and fully observed outcome, $\blacksquare$ ``large'' true CACE and fully observed outcome,  \textcolor{mygray}{$\square$} ``small'' true CACE but partially observed outcome, and $\square$ ``large'' true CACE but partially observed outcome. The dotted line in the bias plot represents no bias. Dashed lines on the coverage plot are the 94 and 96 \% coverage rates.
The CI width resulting from TSRI, where $N=200$, $Y$ has missing data and CACE=4 in logit scale is not plotted as it was $>2000$.}
\label{fig:binary2}
\begin{center}
\includegraphics[scale=.9]{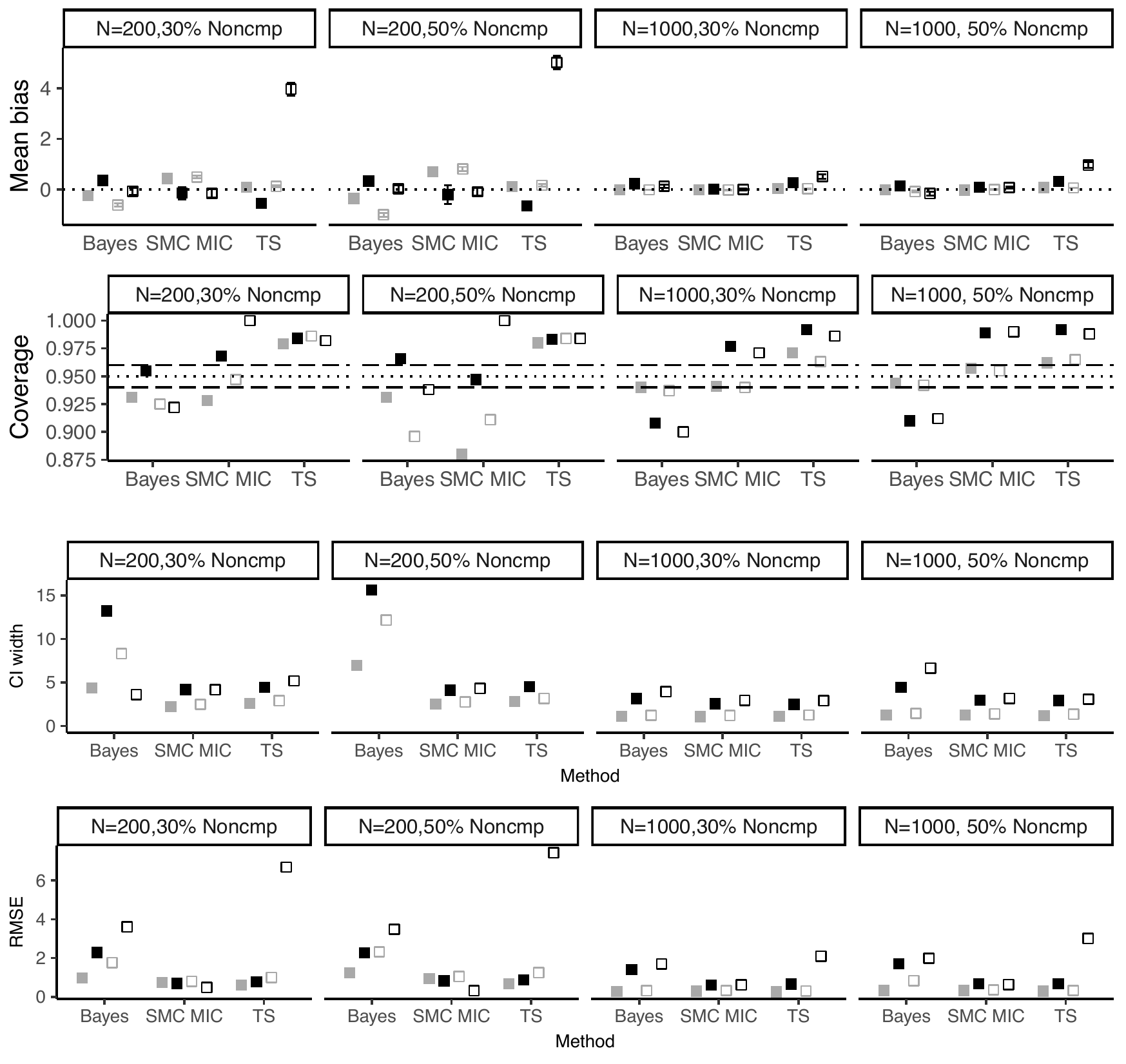}
\end{center}
\end{figure}

Figure \ref{fig:cont} shows the mean bias  (top) and CI coverage rate  and width (middle) and RMSE (bottom) corresponding to scenarios with  continuous outcome. Results  for binary outcomes appear in Figure  \ref{fig:binary} (for $\beta_c=0$) and Figure \ref{fig:binary2} (for $\beta_c=\beta_{cz}/2$).

With continuous outcome,  all methods show close to zero bias, whether the true CACE is small or large, or the outcome is fully observed or partially missing. However, the bias is more pronounced with wider MCE CIs, that do not cross 0 in many settings with small  sample size ($N=200$). This bias is much reduced at the larger sample size,  with MCE CIs that contain 0 in all settings with fully observed outcome when analysing with Bayesian or SMC MIC.  TSLS  exhibits some significant non-zero bias, especially where there are  missing outcomes. 

Coverage rates are close to the nominal value (between 94 and 96\%). CI width is  smaller for SMC MIC and Bayesian methods than TSLS, with SMC MIC slightly outperforming Bayesian methods in many scenarios. 
In terms of  RMSE, we observe that both Bayesian and SMC MIC result in very similar values,  inline with theoretical results about their equivalence. TSLS has a larger RMSE.  These results support the claim that the SMC MIC method is  more efficient than TSLS. 

With binary outcomes, in scenarios where $\beta_c=0$,  Bayesian estimates are more varied, with some significant negative bias where the local odds ratio is large, and the sample size is small (the largest being  13\% in absolute value). All bias becomes negligible at larger sample sizes ($N=1000$). The bias has implications for the coverage in the small sample scenarios, and Bayesian estimation results in low coverage (88\%). Coverage is  acceptable in most other settings. The bias is likely the result of the informative prior used for the log odds ratio, which puts more probability on smaller odds ratios values.  
In contrast,  SMC MIC results in positive bias with small sample sizes, but appears unbiased in larger samples. With small samples it results in over coverage. At larger sample sizes, CI coverage rates is once again between 94\% and 96\%.

TSRI, which  is expected to be consistent in these settings, results in small bias and valid CI coverage in scenarios where $N=1,000$, but substantial bias (around 30\%) is present when the true conditional local log odds ratio is large and the sample size small. 

When comparing CI width and RMSE, we observed similar patterns to those exhibited with continuous outcomes, namely  SMC MIC slightly outperforming Bayesian methods in many scenarios, which in turn outperforms TS methods.

In scenarios where $\beta_c\neq0$, Figure \ref{fig:binary2}  shows all methods report some bias with small sample sizes. The most extreme bias for Bayesian methods is negative 50\%, while for SMC MIC  is 40\% (for both, this occurred where the true conditional CACE is small, there is 50\% noncompliers and missing outcomes). However, both Bayesian and SMC MIC have nearly no bias  when the sample size is $N=1,000$. In contrast, TSRI reports considerable bias even at the larger sample size, 13\% where $\beta_c=1$ and $24\%$ where $\beta_c=2$,  in line with the results of \citet{Cai2011}. 

Regarding coverage rates, Bayesian methods results in coverage above 90\% in all settings.
With the small sample size, SMC MIC results in under-coverage (88\%) in one scenario (due to the considerable bias present. In contrast,  where there is missing outcome data, it reports coverage rates close to 100\%. Coverage is acceptable for the larger sample size though.  TSRI reports coverage rates reasonably close to the nominal values but the CI width  larger than after the other methods. 
In particular,  where $\beta_c\neq0$ the CI width resulting from TSRI, with small sample size, missing outcome data and large true CACE is $>2000$.

In terms of RMSE, in settings where the effect of compliance in the control group is small ($\beta_c=1$), TSRI has smallest RMSE, despite reporting larger biases. In settings with larger effect of complying for the unexposed ($\beta_c=2$), SMC MIC outperforms the other methods.

\section{Motivating example: the COPERS trial} \label{Sec:COPERS}
We now illustrate the methods in practice by applying each in turn to a real RCT. 
The COping with persistent Pain, Effectiveness Research in Self-management trial (COPERS)  was a randomised controlled  trial across 27 general practices and community services in the UK. It recruited 703 adults with musculoskeletal pain of at least three months duration, and randomised 403 participants  to  the active intervention and a further 300 to the control arm. The mean age of participants was 59.9 years, with 81\% white, 67\% female, 23\% employed, 85\% with pain for at least three years, and 23\% on strong opioids.

Intervention participants were offered 24 sessions introducing them to cognitive behavioural (CB) approaches designed to promote self-management of chronic back pain.  The sessions were delivered over three days within the same week with a follow-up session two weeks later.  At the end of the three-day course, participants received a relaxation CD and self-help booklet. Controls received usual care and the same relaxation CD and self-help booklet. The control arm participants had no access to the active intervention sessions. Participants and group facilitators  were not masked to the study arm they belonged to. 

The primary outcome was pain-related disability at 12 months measured by the  Chronic Pain Grade (CPG) disability sub-scale. This is a continuous measure on a scale from 0 to 100, with higher scores indicating worse pain-related disability.  Secondary outcomes included  depression (Hospital Anxiety and Depression Scale (HADS) depression sub-scale, ranging from 0 to 21) and social integration (Health Education Impact Questionnaire (HEIQ) social integration and support sub-scale, range 4--20), amongst others.

The ITT analysis found no evidence that the COPERS intervention had an effect on improving pain-related disability at 12 months ($-1.0$, $95$\% CI $-4.8$ to $2.7$). 

However,  only  179  (45\%) of those randomised to the active treatment attended all 24 sessions, with 322 (86.1\%) receiving at least one session.  Since poor attendance to the sessions was anticipated, the original statistical analysis plan included obtaining the CACE  for primary and secondary outcomes,  adjusted for all of the  baseline covariates included in the primary analysis models, namely site of recruitment, age, gender and HADS depression score at baseline, and the corresponding outcome at baseline. 

The original publication defined those  attending at least half of the course (i.e. those present for at least 12 of the 24 course components) as the compliers, and used TSLS estimation after multiple imputation to estimate the CACE.
It reported no evidence of  a causal treatment effect on CPG disability at 12 months amongst the compliers ( $-1.0$, $95$\% CI $-5.9$ to $3.9$). In contrast, there was evidence supporting a non-zero CACE for two secondary outcomes, depression and social integration  \citep{Taylor2016HTA}. 

For this reason, we focus our re-analyses on these two outcomes. To exemplify the methods,  HEIQ social integration at 12 months is considered as continuous, but HADS depression score is dichotomised by classifying those with a score of 11 and over as depressed, and not depressed otherwise \citep{Bjelland2002}. We analyse each outcome with a substantive model that adjusts for site of recruitment, age, gender and HADS depression score at baseline, and the corresponding outcome at baseline. 
In addition to these variables, our dataset  also contains: Housing (Living arrangements: living alone versus living with others),  whether they speak fluent English, ethnicity,  employment (dichotomised  as employed or in full time education versus not employed or in full-time education), age when left education (categorical), and the baseline measurements of other secondary outcomes, namely  Pain Self-Efficacy Questionnaire (PSEQ), HADS anxiety score, and  Chronic Pain Acceptance Questionnaire (CPAQ).
Thirty-eight individuals  (19 in each arm) originally recruited were completely lost to follow up, and are excluded from this analysis, as per the original COPERS analysis, leaving a sample size of 665 participants, who were followed up for 12 months, 384 allocated to active treatment, and  281 to the control (93\% of those recruited).   

Table \ref{Tabmiss} reports descriptive statistics and percentage of observations with missing data  by treatment group. Geographical location (Warwick or London), age, gender, ethnicity, employment and educational attainment were all fully observed in the 665 participants.  

We re-define treatment received, and consequently compliance, as taking \textit{any} session in the intervention, in order to avoid more obvious exclusion restriction violations.
For those participants completely unexposed to the training sessions,  the exclusion restriction seems plausible, as it is unlikely that that random allocation has a direct effect, but since participants were not blinded to their allocation, we cannot completely rule out  some  psychological effects  of knowing which group they belong to  on depression and disability. 
The other two core IV assumptions, unconfoundedness and instrument relevance are justified by design, as is the monotonicity assumption, since the intervention was not available outside the trial.

We begin by performing a separate MI to analyse the ITT and the TSLS or TSRI respectively.
We use MI by FCS to impute the outcome and the baseline variables with missing data (HADS depression and HEIQ social integration). We use employment, CPG disability, HADS anxiety and CPAQ as auxiliary variables (and HEIQ social integration at baseline for the HADS depression outcome). Since only employment was fully observed, all others also needed to be imputed.
 The Bayesian and SMC MIC methods deal with  the missing data within the procedure, using the same auxiliary variables.

We report the ITT first, and then obtain a CACE with Bayesian, SMC MIC and two-stage methods in Table \ref{Tab:CopersRes}.
As was the case in simulations, Bayesian and SMC MIC estimation methods resulted in very similar estimates, but SMC MIC reports shorter confidence intervals. In the case of continuous outcome, social integration, TSLS also reports a very similar point estimates, but with a larger SE. 

For the binary outcome, HADS depression, we first notice that the point estimate obtained after TSRI is somewhat smaller in absolute value, so the treatment effect is less favourable. In addition the SE is considerably larger than that after either Bayesian or SMC MIC.  As a result, we do not have a significant effect of treatment on the local log odds ratio of being depressed when receiving the COPERS intervention for the compliers, even though the point estimate is larger in absolute value than the ITT. 
 Recall that TSRI requires that there is no effect of being a complier in the  unexposed. It is very plausible that this assumption is violated in the COPERS trial, as the self-help booklet and relaxation CD may have a positive effect in reducing depression, in those that comply.
 
From this reanalysis, we conclude that there is a small positive causal effect of the COPERS intervention on improving social integration and reducing depression in those that comply by attending at least one session. These results depend on random allocation to treatment being a valid instrument, which as discussed previously seems a reasonable assumption. 
The local causal effects found are very small, and possibly not clinically relevant. This may be the result of our very low ``dose'' definition of compliance, which is however necessary to avoid violations of the exclusion restriction. 

\section{Discussion}\label{Sec:Discussion} 
This paper proposes the use of SMC MI of latent compliance classes  to estimate the CACE using fully parametric mixture models fitted by maximum likelihood, and combined via Rubin's rules. 
 We have demonstrated empirically through simulations that the SMC MIC estimation has good finite sample performance, which is  approximately equal to Bayesian estimation, and compares favourably to two-stage methods especially for estimating causal odds ratios. 

The efficiency gains (in terms of SEs and narrower CIs)  are more pronounced when auxiliary variables are incorporated. This is easily done within SMI MIC, because MI separates the imputation and the analysis stages, making it possible to  estimate  marginal local effects, i.e. only conditional on compliance class and treatment received. This is especially important for local odds ratios, given their non-collapsibility.   
Moreover, we have shown that for the estimation of local odds ratio, two-stage methods are only valid with certain modifications, for example by including the residuals from the first stage into the second stage and then again only in some special cases, namely where there is no effect of complying on the unexposed.  While Bayesian estimation of the mixture models is valid, it is sensitive to the choice of priors, especially in small sample settings, and inclusion of several auxiliary variables requires careful coding, and the use of specialist software, making SMC MIC  preferable to use in practical applications.

\citet{Taylor2009} previously proposed using MI for imputing the compliance classes in  settings with missing outcomes, and compared MI estimation to existing Bayesian and frequentist methods. However, this work used a made-to-measure sampler, thus limiting its use in practice. In contrast, our method  uses MI procedures already available in widely used statistical software packages (R and Stata).  Another difference of our contribution is the focus on estimation where the estimand of interest for a binary outcome is the local causal odds ratios. 

While the methods were exemplified and tested in the context of  estimating the LATE in RCTs with non-adherence, binary exposure and binary compliance classes,  assuming no always-takers and no defiers, SMI MIC is easily applicable to settings with always-takers, by simply using an multinomial imputation model for the compliance class, already implemented in the smcfcs packages. Also, although  we assumed the strong exclusion restriction, it is possible to relax this, so that it only needs to hold for either the always-takers  or the never-takers \citep{Hirano2000}. As our analysis of the illustrative example shows, the models  can easily be modified to include baseline covariates  \citep{Little1998,Hirano2000}. Moreover, this makes it possible to apply these  methods to situations  where the IV assumptions are  only satisfied after conditioning on other variables, thus extending the applicability to certain observational settings where conditional IVs are more plausible. 

The present study has some limitations.  We have focused on the LATE estimand, which is often criticised because the estimates obtained apply to a stratum of the population  (the compliers) which cannot be observed in practice, thus limiting its applicability. However, the LATE may provide some useful information about the average causal effect in the entire population. See Baiocchi \citep{Baiocchi2014} for a discussion on this. 
Moreover, the average treatment effect on the compliers may be of interest in its own right, and patients are interested in this being estimated and reported,  especially  when they expect  to comply with the treatment \citep{Murray2018}. In such cases, it may be  desirable to describe the compliers, by modelling their distribution in terms of observed characteristics \citep{Brookhart2006, AngristPischke2009}. 

All methods considered here assume a specific parametric model and as such the estimates are sensitive to these parametric assumptions \citep{Tan2006}.
In addition, we consider only situations where  the identification assumptions hold.
There are several options to study the sensitivity to departures from these  assumptions.  For example, if the exclusion restriction does not hold, a Bayesian parametric model can use priors on the non-zero direct effect of  randomisation on the outcome for identification \citep{Conley2012}. Since these models are only weakly identified,  the results depend strongly on the  prior distributions.
Alternatively, violations of the exclusion restriction can also be handled by using baseline covariates to model the probability of compliance directly, within structural equation modelling via expectation maximisation \citep{Jo2002, Jo2002a}.

Settings where the instrument is only weakly correlated with the exposure have not been considered here. Nevertheless, Bayesian methods have been shown to perform better than TS methods when the instrument is weak  \citep{Burgess2012}.
 Weak instruments may result in larger biases, which may be present due to model misspecification. Since the methods presented are not robust to such misspecifications, this is a particular concern.  
 This suggest a possible extension to the work presented here, where more flexible non-parametric MI could be of use.  

Extending our proposal to hierarchical settings,  for example for cluster randomised trials with non-compliance, where the mixture models  include a random effect for clustering, should be straightforward, as the compliance classes in these models can be imputed in a substantive model compatible manner  by using the newly developed SMC functionality of the multilevel MI R package JOMO \citep{Quartagno2018}. However, careful consideration of the  assumptions in the presence of (partial) interference  at the cluster level is required \citep{Sobel2006}.
Future research directions could include exploiting the additional flexibility of SMC MIC and extend our method  to handle time-varying non-compliance in longitudinal settings.

In summary, our results show that SMC MIC provides a practical inferentially reliable method for estimation of the local average treatment effects, even when the target estimand is a causal odds ratio, and in the presence of missing outcome data.

\section*{Acknowledgements}

The authors thank Prof. Stephanie Taylor and the COPERS trial team  for access to the data. 

Karla DiazOrdaz is supported by UK  Medical Research Council Skills Development Fellowship MR/L011964/1.
James Carpenter is supported by the Medical Research Council, grant numbers MC UU 12023/21 and MC UU 12023/29.

\vspace*{-8pt}


\section*{Supplementary Materials}
Web Appendix A, referenced in Section  \ref{Sec:MI-C} and 
Web Appendix B, referenced in Section  \ref{Sec:Simulation} are available with
this paper at the website on Wiley Online
Library.

\clearpage
\begin{table}
  \centering
  \caption{Simulation factors and levels}\label{simulation_factors}
    \begin{tabular}{lll}\hline
    Factor &  \multicolumn{2}{l}{Levels}       \\ \hline
    sample size $N$ & {200} & {1000}   \\
    percentage of compliers &70 & 50   \\
    type of outcome & Continuous & Binary  \\
    conditional true CACE in the link scale  $\beta_{cz}$& $2$ & $4$ \\
   causal effect of complying on the unexposed$^{a}$ $\beta_c$ & $0$ & $\beta_{cz}/2$   \\
    missing outcome & no    & $20\%$ MAR given $X_2$  \\ \hline
 \hline
\multicolumn{3}{l}{$^{a}$ \scriptsize{only for binary outcome scenarios}} \\
\hline
    \end{tabular}%
  \label{sim:factors}%
\end{table}%

\begin{table}\centering
\caption{The COPERs study: descriptive statistics and description of missing data, by treatment group.}
\label{Tabmiss} 
\begin{tabular}{lll}\hline \hline
          & Control & Intervention \\ \hline\hline
N analysed (N=665)&$281 \  (42)$ & $384 \ (58)$\\
    N (\%) Switched & $0$ $(0)$ & $53$  $(14)$ \\
   \hline
   \hline
\small{\it Fully observed Baseline variables}\\\hline
\hline
age  \\
Mean (SD)  & $59.64\  (13.38)$ & $60.37\  (13.48)$ \\
   \hline
gender \\
Male (\%)  & $90\  (32.0)$ & $128\  (33.3)$ \\
   \hline
site of recruitment \\
Warwick (\%)  & $154\  (0.55)$ & $ 215\  (0.56)$ \\
   \hline
   \hline
Employment \\
full time  (\%)  & $72  \  (0.26)$ & $ 88\  (0.23)$ \\
\hline
\hline
    \small{\it Baseline variables with missing data} &   & \\ \hline\hline
CPG disability\\
N (\%) missing & $3$ $(<1)$ & $4$  $(1)$ \\
Observed Mean (SD) & $50.85\  (19.15)$ & $ 50.46 \ (18.99)$ \\
\hline
HEIQ social integration\\
N (\%) missing & $5$ $(2)$ & $2$  $(<1)$ \\
Observed Mean (SD) & $13.91 \ (3.38)$ & $13.97\ (3.57)$ \\
   \hline
HADS depression score\\
N (\%) missing & $3$ $(<1)$ & $2$  $(<1)$ \\
Observed Mean (SD) & $7.32\ (3.91) $ & $7.41 \ (4.13)$ \\
 \hline
HADS anxiety\\
N (\%) missing & $3$ $(<1)$ & $3$  $(<1)$ \\
Observed Mean (SD) & $9.20\ (4.70) $ & $9.27\ (4.56) $ \\
   \hline
\hline
\small{\it Outcomes at 12 months} \\ \hline\hline
HEIQ social integration\\
N (\%) missing & $40$ $(14.2)$ & $34$  $(8.9)$ \\
Observed Mean (SD) & $ 14.06\  (3.60)$ & $ 14.89 \ (3.54)$ \\
\hline
HADS depression score\\
N (\%) missing & $35$ $(12.5)$ & $37$  $(9.6)$ \\
Observed Mean (SD) & $ 6.94\  (4.61)$ & $ 6.16\ (4.33)$ \\
\hline \hline 
  \end{tabular}%
\end{table}

\begin{table}\centering
\caption{The COPERs study: CACE based on the available cases (665 participants).}
\label{Tab:CopersRes} 
\begin{tabular}{llrr}\hline \hline
\quad Method & \quad CACE & SE & 95\% CI \\
\hline
HEIQ social integration\\
\quad ITT & \quad $0.77$ & $0.21$&  $(0.35, 1.19)$\\
\quad Bayesian &\quad $0.86$&  $0.24$ &$(0.37, 1.33)$\\ 
\quad SMC MIC&\quad $0.88$ & $0.24$ &$(0.41,1.35)$\\
\quad TSLS &\quad  $0.85$& $0.25$&  $(0.38, 1.33)$\\
\hline

HADS depression &(log-odds ratio scale)\\
\quad ITT & $-0.59$ & $0.23$&  $(-1.05, -0.12)$\\
\quad Bayesian &$-0.74$&  $0.27$ &$(-1.27, -0.21)$\\ 
\quad SMC MIC&$-0.73$ & $0.27$ &$(-1.27,-0.19)$\\
\quad TSRI & $-0.68$& $0.36$&$(-1.38, \ \ 0.02)$\\
\hline \hline 
  \end{tabular}%
\end{table}

\end{document}